\documentclass{astlb}

\usepackage[hyperfootnotes=false]{hyperref}

\usepackage{graphicx}

\usepackage{url}

\renewcommand{\deg}{\ensuremath{^\circ}}

\def\Msolar{\ifmmode{\rm M}_{\mathord\odot}\else${\rm M}_{\mathord\odot}$\fi}

\def\,{\thinspace}
\def\lsim{\mathrel{\raise .4ex\hbox{\rlap{$<$}\lower 1.2ex\hbox{$\sim$}}}}
\def\gsim{\mathrel{\raise .4ex\hbox{\rlap{$>$}\lower 1.2ex\hbox{$\sim$}}}}

\def\simprop{\mathrel{\raise .4ex\hbox{\rlap{$\propto$}\lower 1.2ex\hbox{$\sim$}}}}
\def\deg{\ifmmode^\circ\else$^\circ$\fi}
\def\pdeg{\ifmmode $\setbox0=\hbox{$^{\circ}$}\rlap{\hskip.11\wd0 .}$^{\circ}
          \else \setbox0=\hbox{$^{\circ}$}\rlap{\hskip.11\wd0 .}$^{\circ}$\fi}
\def\arcs{\ifmmode {^{\scriptstyle\prime\prime}}
          \else $^{\scriptstyle\prime\prime}$\fi}
\def\arcm{\ifmmode {^{\scriptstyle\prime}}
          \else $^{\scriptstyle\prime}$\fi}
\newdimen\sa  \newdimen\sb
\def\parcs{\sa=.07em \sb=.03em
     \ifmmode \hbox{\rlap{.}}^{\scriptstyle\prime\kern -\sb\prime}\hbox{\kern -\sa}
     \else \rlap{.}$^{\scriptstyle\prime\kern -\sb\prime}$\kern -\sa\fi}
\def\parcm{\sa=.08em \sb=.03em
     \ifmmode \hbox{\rlap{.}\kern\sa}^{\scriptstyle\prime}\hbox{\kern-\sb}
     \else \rlap{.}\kern\sa$^{\scriptstyle\prime}$\kern-\sb\fi}
\def\ra[#1 #2 #3.#4]{#1\sup{h}#2\sup{m}#3\sup{s}\llap.#4}
\def\dec[#1 #2 #3.#4]{#1\deg#2\arcm#3\arcs\llap.#4}
\def\deco[#1 #2 #3]{#1\deg#2\arcm#3\arcs}
\def\rra[#1 #2]{#1\sup{h}#2\sup{m}}
\def\doubleline{\vskip 3pt\hrule \vskip 1.5pt \hrule \vskip 5pt}

\begin{document}

\journalinfo{2016}{42}{2}{63}[68]

\title{Additional spectroscopic redshift measurements for galaxy clusters from
the First Planck Catalogue}

\author{\bf 
V.S. Vorobyev\address{1},
R.A. Burenin\email{rodion@hea.iki.rssi.ru}\address{1},
I.F.Bikmaev\address{2,3},
I.M.Khamitov\address{2,4},\\
S.N. Dodonov\address{5},
R.Ya. Zhuchkov\address{2,3},
E.N. Irtuganov\address{2,3},
A.V.Mescheryakov\address{1,2},\\
S.S. Melynikov\address{2,3},
A.N. Semena\address{1},
A.Yu. Tkachenko\address{1},
N. Aghanim\address{6},
R.~Sunyaev\address{1,7}
\addresstext{1}{Space Research Institute of the Russian Academy of Sciences, Moscow, Russia}
\addresstext{2}{Kazan Federal University, Kazan, Russia}
\addresstext{3}{Academy of Sciences of The Republic of Tatarstan, Kazan, Russia}
\addresstext{4}{Scientific and Technological Research Council of Turkey, Antalya, Turkey}
\addresstext{5}{The Special Astrophysical Observatory of the Russian Academy of Sciences, Nizhnij Arkhyz, Russia}
\addresstext{6}{Institut d\'Astrophysique Spatiale, Orsay, France}
\addresstext{7}{Max Planck Institute for Astrophysics, Garching, Germany}
} 

\shortauthor{VOROBYEV et al.} 
\shorttitle{ADDITIONAL SPECTROSCOPIC REDSHIFT MEASUREMENTS}

\submitted{September 3, 2015} 

\begin{abstract}
  We present the results of spectroscopic redshift measurements for
  the galaxy clusters from the first all-sky Planck catalogue of the
  Sunyaev-Zeldovich sources, that have been mostly identified by means
  of the optical observations performed previously by our team
  \citep{Pl_RTT150}. The data on 13 galaxy clusters at redshifts from
  $z$ $\approx$ $0.2$ to $z$ $\approx$ $0.8$, including the improved
  identification and redshift measurement for the cluster
  PSZ1\,G141.73+14.22 at $z = 0.828$, are provided. The measurements
  were done using the data from Russian–Turkish 1.5-m telescope
  (RTT-150), 2.2-m Calar Alto Observatory telescope, and 6-m SAO RAS
  telescope (Bolshoy Teleskop Azimutalnyi, BTA).


\end{abstract}

\section{Introduction}

A large number ($\sim10^3$) of galaxy clusters are detected in the
Planck all-sky survey \citep{PSZcat13,PSZ1Addendum,PSZ2} via the
Sunyaev-Zeldovich (SZ) effect \citep{SZ}. The sensitivity of the
survey turns out to be distributed nearly more or less uniformly over
the entire sky. Since the SZ signal amplitude depends mainly on the
mass of clusters and not on their redshift, all the most massive (with
masses $>6\cdot 10^{14}$$\Msolar$)
galaxy clusters in the Universe are detected in the Planck
survey. This sample of galaxy clusters is unique and very important
for various cosmological studies, such as constraining cosmological
parameters using the measurements of the galaxy cluster mass function
\cite[see, e.g.,][]{cccp,PSZcosm13,PSZ2cosm}.

Many of the clusters detected in the \emph{Planck} survey are known
galaxy clusters that were detected previously in various optical or
X-ray surveys. However, about half of the detected objects turn out to
be previously unknown clusters. For these objects additional optical
observations should be carried out in many cases, to optically
identify them with galaxy clusters and to measure their
redshifts. This work is carried out at many telescopes
\citep{PSZcat13, Pl_RTT150, PSZ1Addendum, PSZ2, Canary}; our team also
participates in this work.\tabularnewline

For some of the clusters identified by our team previously, only
photometric redshift estimates were reported \citep{Pl_RTT150}. The
accuracy of these estimates (about 3\%) is insufficient for an
accurate measurement of the cluster mass function. To make
spectroscopic redshift measurements for these clusters, we carried out
additional optical observations at RTT-150, the 2.2-m Calar Alto
Observatory telescope, and the 6-m BTA telescope during 2014. The
results of these measurements are presented below.

\begin{figure*}
  \centering
  \includegraphics[width=0.8\linewidth]{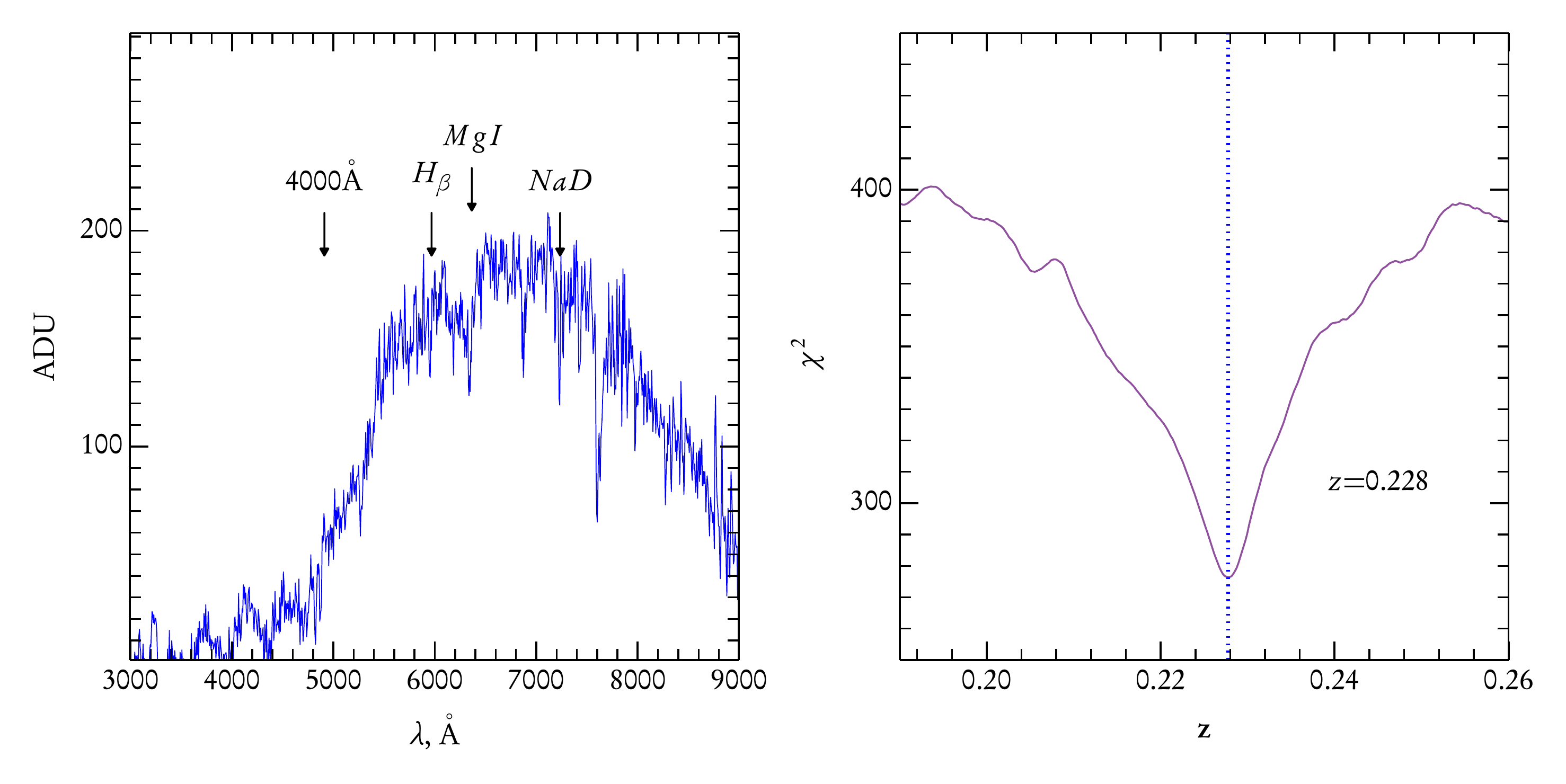}
  \caption{The spectrum of the brightest galaxy of cluster
    PSZ1\,G114.81$-$11.80, $z = 0.2277$, obtained with TFOSC
    spectrometer at RTT-150 telescope (left) and $\chi^2$ from the
    cross-correlation with an elliptical galaxy template spectrum
    (right).}
  \label{fig:rtt_example}
\end{figure*}

\begin{figure*}
  \centering
  \includegraphics[width=0.8\linewidth]{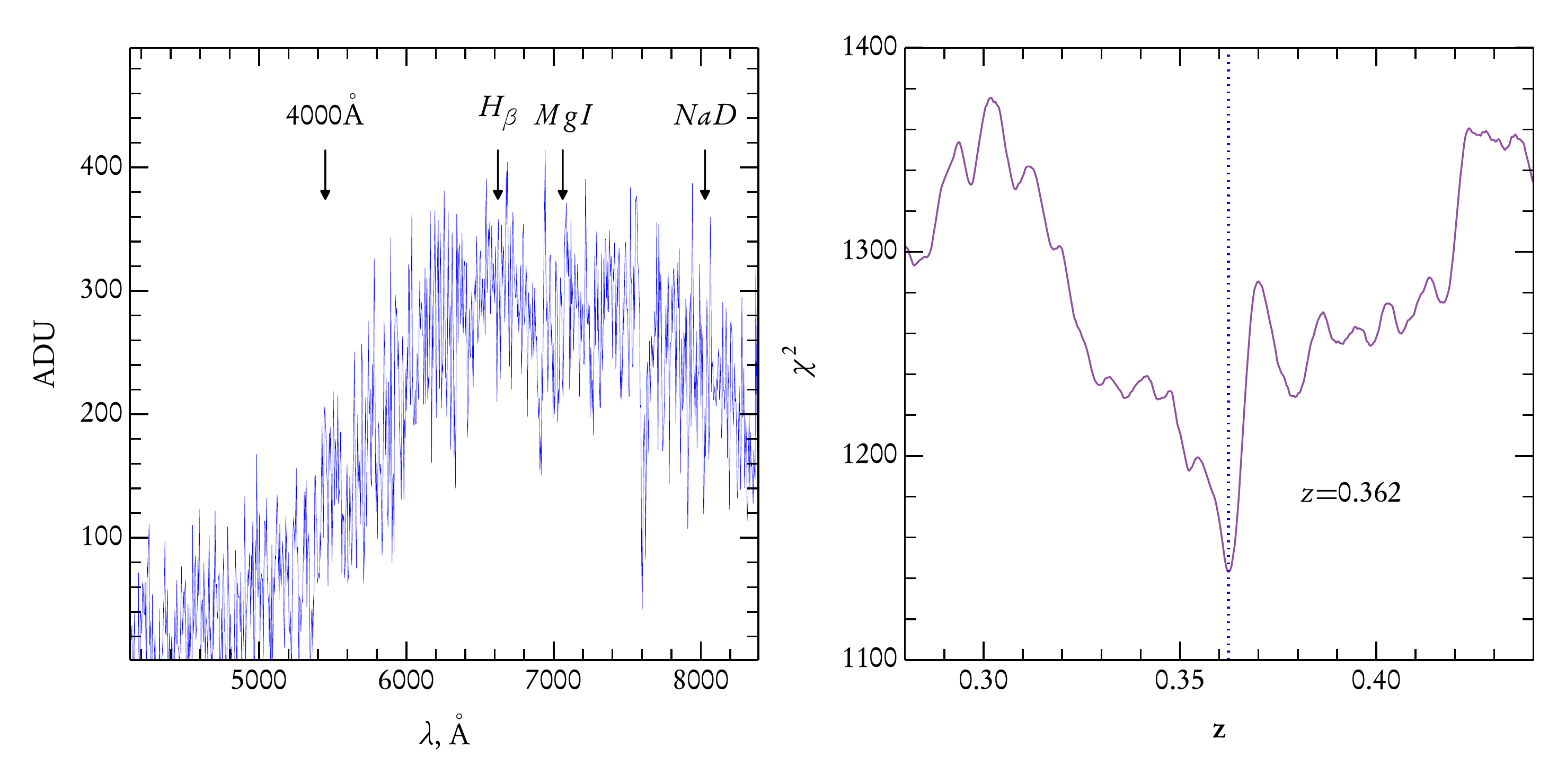}
  \caption{The spectrum of the brightest galaxy of cluster
    PSZ1\,G157.84$+$21.23, $z = 0.3623$, obtained with CAFOS
    spectrometer at  2.2-m Calar Alto
    Observatory telescope (left) and $\chi^2$ from the
    cross-correlation with an elliptical galaxy template spectrum
    (right).}
  \label{fig:caha2.2m_example}
\end{figure*}

\begin{figure*}
  \centering
  \includegraphics[width=0.8\linewidth]{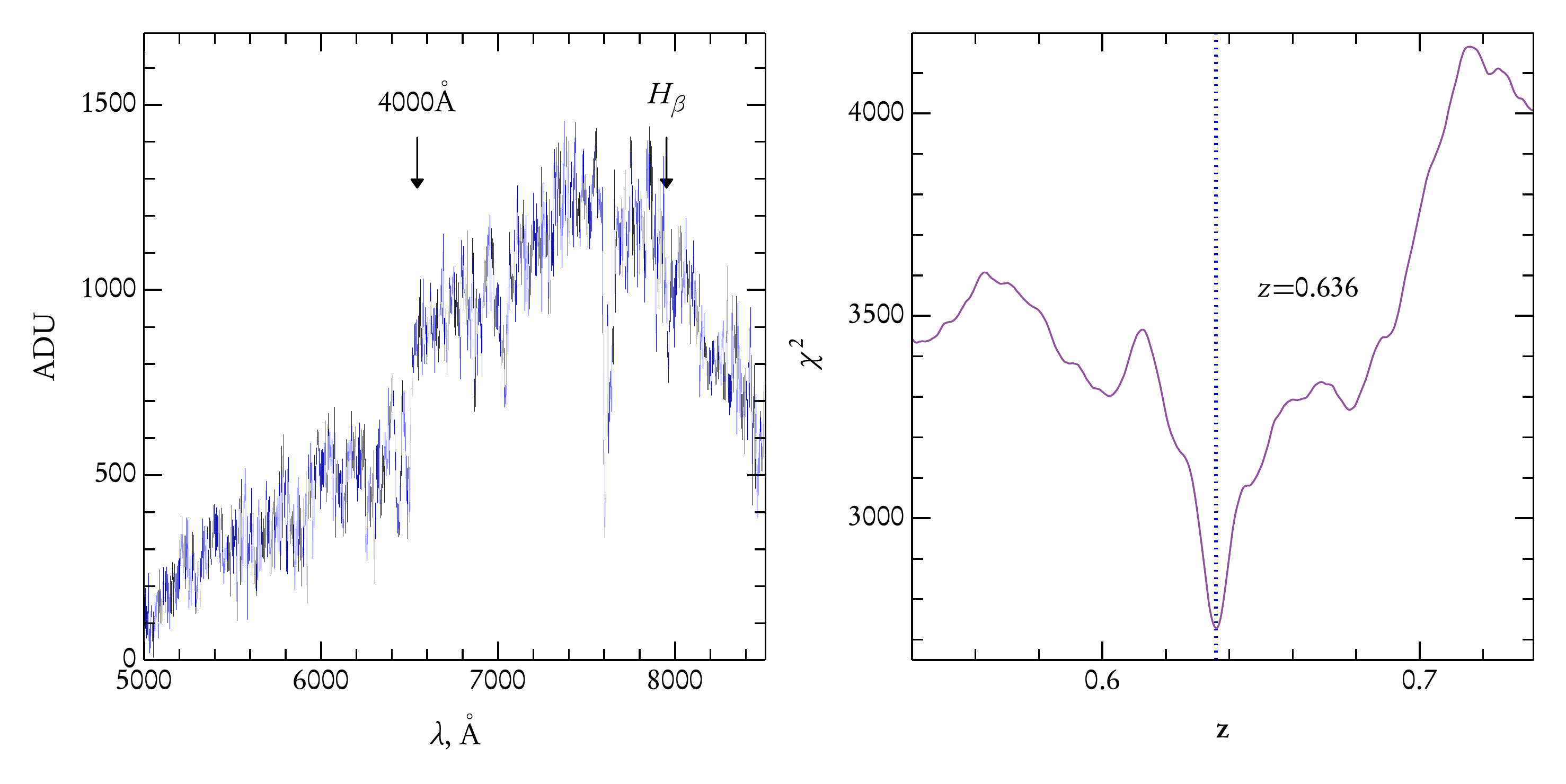}
  \caption{ The spectrum of brightest cluster's galaxy PSZ1\,
    G183.26$+$12.25, $z = 0.6359$, obtained with the 6-m BTA telescope
    using the \emph{SCORPIO-2} spectrometer (left) and $\chi^2$ from
    the cross-correlation with an elliptical galaxy template spectrum
    (right).  }
  \label{fig:bta_example}
\end{figure*}

\begin{figure*}
  \centering
  \includegraphics[width=0.8\linewidth]{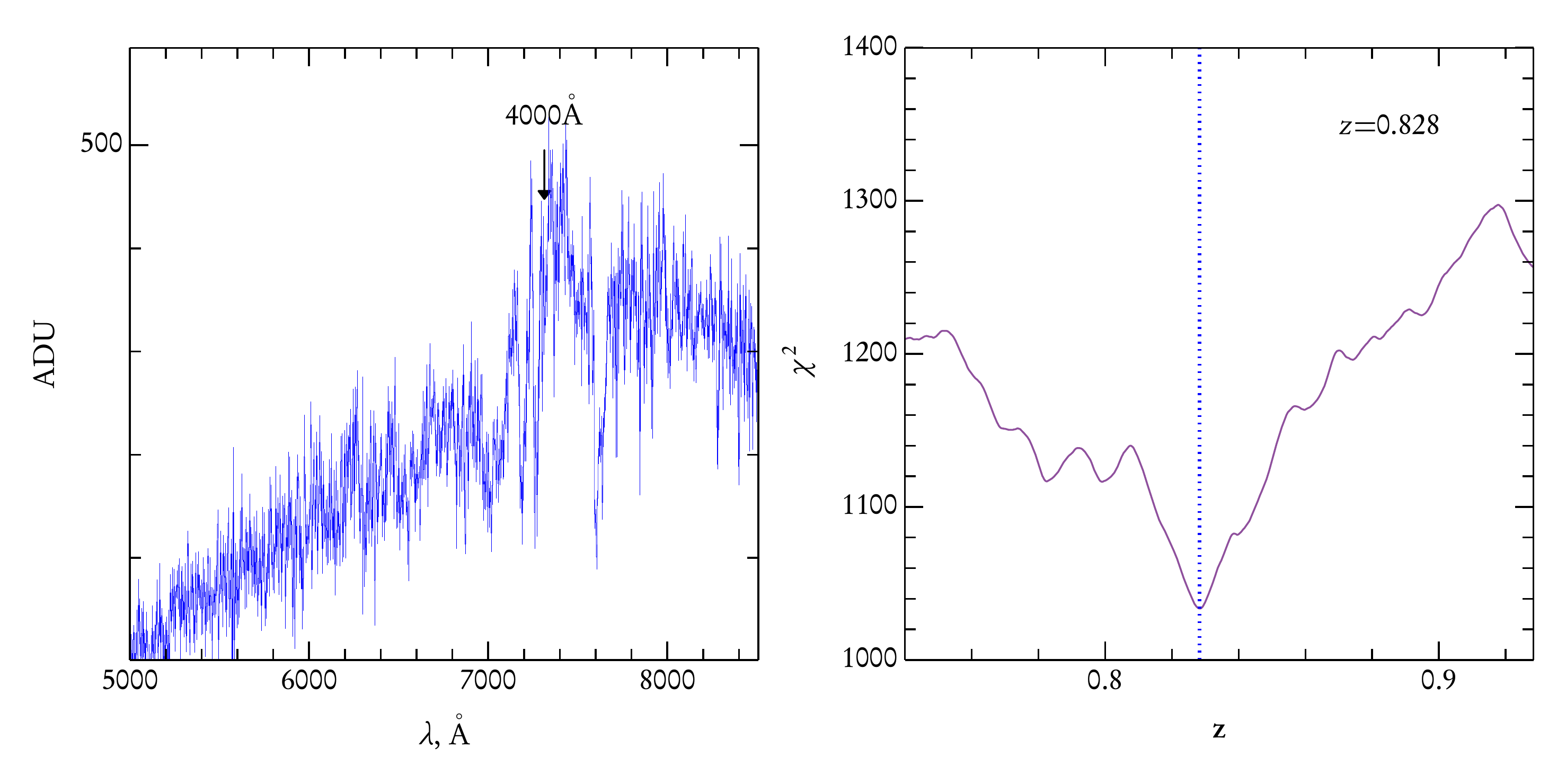}
  \caption{The spectrum of brightest cluster's galaxy
    PSZ1\,G141.72+14.22, $z = 0.8283$, obtained with the 6-m BTA
    telescope using the \emph{SCORPIO-2} spectrometer (left), and
    $\chi^2$ from the cross-correlation with an elliptical galaxy
    template spectrum (right).}
  \label{fig:G14172_spec}
\end{figure*}

\begin{table*}
  \caption{Redshifts of galaxy clusters} 
  \label{tab:clres}
  \vskip 2mm
  \renewcommand{\arraystretch}{1.1}
  \renewcommand{\tabcolsep}{0.35cm}
  \centering
  \footnotesize
  \begin{tabular}{lcccll}
    \noalign{\doubleline}
    & \multispan2\hfil Coordinates (J2000)\hfil\\
    Name & $\alpha$ & $\delta$ & $z$ & telescope$^*$ & Other name \\
    \noalign{\vskip 3pt\hrule\vskip 5pt}
    PSZ1 G048.22$-$65.03 & 23 09 51.0 & $-$18 19 57 & 0.413 & 2,3\\
    PSZ1 G060.12$+$11.42 & 18 58 46.0 & $+$29 15 34 & 0.225 & 1\\
    PSZ1 G071.57$-$37.96 & 22 17 15.8 & $+$09 03 10 & 0.291 & 1 & ACO 2429\\ 
    PSZ1 G080.11$-$77.29 & 00 15 24.4 & $-$17 30 34 & 0.462 & 3\\
    PSZ1 G134.31$-$06.57 & 02 10 25.1 & $+$54 34 09 & 0.334 & 3\\ 
    PSZ1 G141.73$+$14.22 & 04 41 05.8 & $+$68 13 16 & 0.828 & 3\\
    PSZ1 G157.44$+$30.34 & 07 48 54.3 & $+$59 42 06 & 0.403 & 2 & [ATZ98] B100 \\ 
    PSZ1 G157.84$+$21.23 & 06 40 32.7 & $+$57 45 36 & 0.363 & 2,3\\
    PSZ1 G183.26$+$12.25 & 06 43 09.9 & $+$31 50 55 & 0.636 & 3\\ 
    PSZ1 G205.56$-$55.75 & 03 15 22.0 & $-$18 12 22 & 0.236 & 1\\ 
    PSZ1 G210.55$-$44.61 & 04 03 42.5 & $-$17 08 04 & 0.143 & 2 & ACO 472\\ 
    PSZ1 G223.04$-$20.27 & 05 54 37.3 & $-$17 44 35 & 0.163 & 1\\
    PSZ1 G224.01$-$11.14 & 06 30 55.3 & $-$14 51 00 & 0.560 & 3\\
    \noalign{\vskip 3pt\hrule\vskip 5pt}
  \end{tabular}

  \begin{minipage}{0.8\linewidth}
    \parindent 0mm
    
    The telescope at which the cluster redshift was measured: 1 for
    the 1.5-m RTT-150 telescope; 2 for the 2.2-m Calar Alto
    Observatory telescope; 3 for the 6-m BTA telescope.
  \end{minipage}
\end{table*}

\section{Observations} 

Cluster member galaxies are identified trough the observation of the
red sequence of galaxies in the color-magnitude diagram. The
photometric redshift estimates for clusters were previously obtained
from the red sequence colors \citep{Pl_RTT150}. To obtain a reliable
cluster spectroscopic redshift measurement, it is sufficient to
measure the redshift of several brightest cluster member galaxies at
the cluster center, or even one brightest cluster galaxy at the center
of a regular cluster. Spectroscopic redshifts can be efficiently
measured for galaxy clusters at $z<0.4$ with \mbox{1.5--2-m}
telescopes. Larger telescopes should be used to measure the redshifts
of clusters at higher $z$.

The procedure of cluster optical identifications and measuring their
redshifts used in our work are based on those developed for
\emph{400d} X-ray survey of galaxy clusters \citep{400d} and for
\emph{160d} survey \citep{160d} earlier. The data of WISE IR-survey
\citep{wright10} were used to search for distant galaxy clusters among
unidentified SZ sources, as described in \cite{clwise}. This procedure
is described in more detail in \cite{Pl_RTT150}.

Optical spectra of galaxy clusters were obtained with the RTT-150
telescope using the \emph{TFOSC} (T\"{U}B\.ITAK Faint Object
Spectrograph and
Camera\footnote{\url{http://hea.iki.rssi.ru/rtt150/}}) spectrograph in
longslit mode. We used grism (\#15) with $\approx$12\AA\ resolution in
3900--9100\AA\ band, and 100~$\mu$m (1.8\arcs) size
slit. Spectroscopic observations presented in this paper, were done
during seven nights in the end of 2013 and during 2014 year at the
telescope RTT-150.

Some redshift measurements of clusters were performed with Calar Alto
Observatory 2.2-m telescope using the CAFOS (Calar Alto Faint Object
Spectrograph)\footnote{\url{http://w3.caha.es/alises/cafos/cafos22.html}}
spectrograph. Observations were carried out during 4 nights in autumn,
2014. For spectroscopic measurements G-200 grism was used, which
provide spectral resolution of about 10\AA\ in 4000--8500 \AA\
spectral band.

The spectroscopic redshift measurements for distant clusters ($z>0.4$)
were made with the 6-m SAO RAS telescope (Bolshoy Teleskop
Azimutal'nyu, BTA) using the \emph{SCORPIO-2} spectrometer
\citep{scorpio_05,scorpio_11}. We used volume phase holographic
grating 940@600, which provide about 10\AA\ spectral resolution in
4000--8500 \AA\ band. To obtain spectroscopic data we used three
observing nights in November, 2014, during which we were able to get
about 12 hours of observational time.

The observations were carried out in approximately the same way at all
telescopes. Typically, a series of two or three spectra was taken for
each slit with an exposure time of 600–1200 s; the spectra of
flat-field and comparison lamps were also taken. During the
subsequent reduction, the series of spectra for the object was aligned
along the spatial axis and combined into a single spectrum for the
subsequent extraction of one-dimensional spectra. All spectroscopic
data were reduced with the standard
\texttt{IRAF}\footnote{\url{http://iraf.noao.edu/}} software package,
which provides the tools to reduce the spectra obtained with long-slit
spectroscopy, and using our own software.

It turns out that to obtain sufficiently accurate spectroscopic
redshift measurements it is not required to get a very high
signal-to-noise ratio spectra of elliptical galaxies. Even if
individual spectral features are not detected, redshift can be
measured accurately from cross-correlation with elliptical galaxy
template spectrum. Accuracy of spectroscopic redshift measurements
presented in this work can be estimated from delta $\chi^2$ and by
comparison with other available redshifts measurements. This accuracy
is not worse than 0.5\%, except for one case where the resulting
spectrum is too noisy and the accuracy of $z$ measurement is about 1\%
(see \ below).

As an example, the spectra of the brightest galaxies in clusters at
various redshifts taken at RTT-150, Calar Alto Observatory 2.2-m
telescope, and BTA 6-m telescope are shown in
Figs.~\ref{fig:rtt_example}--\ref{fig:G14172_spec} (left panel). Right
panels in these Figures show $\chi^2$ from cross-correlation of the
galaxy spectrum with template spectrum of an elliptical galaxy. The
minimum of $\chi^2$ corresponds to the most probable redshift of the
elliptical galaxy.

\section{Results}

The results of our spectroscopic measurements for galaxy clusters from
the first catalogue are presented in Table~\ref{tab:clres}. The object
names, the coordinates of the cluster optical centers, and the
redshifts measured here are provided in this Table. The coordinates of
the cluster optical centers were taken from \cite{Pl_RTT150} and are
given here for completeness. The last two columns specify which
telescope was used to measure the redshift. For two clusters,
PSZ1\,G060.12$+$11.42 and PSZ1\,G071.57$-$37.96, the redshifts were
also measured at the European Northern Observatory telescopes
\citep{Canary}, and these measurements are in good agreement with
ours.

Below, we give comments on several individual objects.

\paragraph{PSZ1\,G141.73$+$14.22}

\begin{figure}
  \centering
  \includegraphics[width=0.8\linewidth]{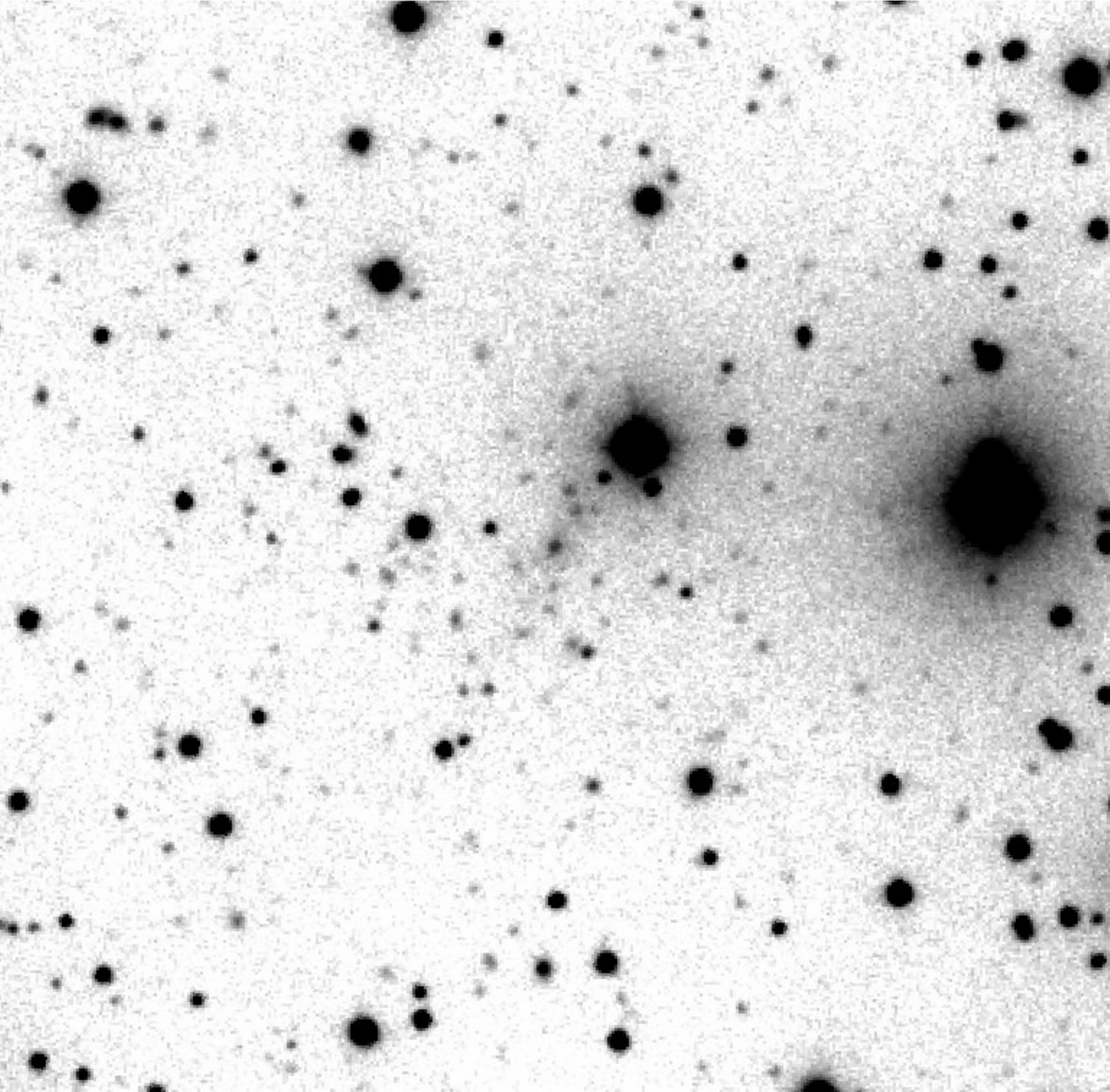}
  \caption{Deep direct image of cluster PSZ1\, G141.72+14.22
    ($z=0.8283$) in \emph{i} filter, obtained at the 2.2-m Calar Alto
    Observatory telescope.}
  \label{fig:G14172_i}
\end{figure}

The measured redshift of this cluster, $z=0.828$, in the table is
considerably more accurate (the error is about 0.2\%) than its
redshift in \cite{Pl_RTT150}. The reason is that in our work we used
the spectrum of the brightest cluster galaxy measured with a
considerably higher signal-to-noise ratio, which was obtained at the
6-m BTA telescope (see Fig.~\ref{fig:G14172_spec}). In addition, using
the 2.2-m Calar Alto Observatory telescope, we were able to obtain a
considerably deeper image of this cluster (see
Fig.~\ref{fig:G14172_i}), which assure the optical identification of
this object.

\paragraph{PSZ1 G157.44$+$30.34}

The spectrum of the brightest galaxy in this cluster taken at the
2.2-m Calar Alto Observatory telescope is noisy. Nevertheless, we
estimate the error in the redshift of this object to be not larger
than 1\%. Larger exposure of the object is needed to make a
spectroscopic measurement with an accuracy considerably better than
1\%.

\paragraph{PSZ1\,G183.26$+$12.25}

The photometric redshift estimate $z_{\mbox{\scriptsize phot.}}= 0.85$
is given in \cite{Pl_RTT150} for this cluster.  The spectroscopic
redshift measured here and given in the table is remarkably different
from the original photometric estimate: $z=0.636$ (see
Fig.~\ref{fig:bta_example}). The reason is that the direct images from
which the photometric $z$ estimate was made were obtained under
uncertain photometric conditions, and, apparently, these conditions
actually appers to be not suitable for the photometric measurements
with the required accuracy.

\section{Conclusion}

At present, the program of optical observations of galaxy clusters
from the first \emph{Planck} Sunyaev-Zeldovitch sources catalogue of
clusters may be considered to be completed. In result of this work,
214 galaxy clusters were detected, and, thus, the first catalogue
contains 947 confirmed galaxy clusters; the spectroscopic redshifts
were measured for 736 of them \citep{PSZ1Addendum}. During 2011--2013,
using observations at RTT150 and the 6-m BTA telescopes, our team
detected 47 previously unknown clusters and measured the spectroscopic
redshifts for 65 clusters \citep{Pl_RTT150}.

In this paper, we present the spectroscopic redshift measurements for
12 more clusters; for one distant cluster (PSZ1\,G141.73$+$14.22,
$z=0.828$), the accuracy of the redshift mesurement was improved
significantly. The corresponding observations were performed during
2014 at RTT150 and 6-m BTA as well as at the 2.2-m Calar Alto
Observatory telescope. Thus, the contribution of our team to the
optical identification and redshift measurement of clusters from the
first \emph{Planck} catalogue is significant.

Recently, the second catalogue of galaxy SZ sources \citep{PSZ2} was
published, which includes 1653 objects, of them 1203 are confirmed
galaxy clusters. Our team will continue the optical observations of
clusters from this catalogue with RTT150, 6-m BTA, the Calar Alto
Observatory telescopes, and, probably, other telescopes.

All clusters from the Planck survey most probably will be detected in
the future Spectrum-R\"oentgen-Gamma (\emph{SRG}) space observatory
all-sky X-ray galaxy cluster survey. Therefore, our observations may
also be considered as the beginning of work on the optical support for
the future SRG/eROSITA survey.

\acknowledgements\tabularnewline
\section{ACKNOWLEDGEMENTS}

This work was supported by the Russian Foundation for Basic Research
(project no. 13-02-12250-ofi-m). We thank T\"UBITAK, the Space
Research Institute, the Kazan Federal University, and the Academy of
Sciences of Tatarstan for partial support in using the Russian–Turkish
1.5-m telescope in Antalya (RTT150). I.F.  Bikmaev, R.Ya. Zhuchkov,
and A.V. Mescheryakov thank the Government of the Russian Federation
for financial support through the subsidy within the framework of the
Program aimed at increasing the competitiveness of the Kazan Federal
University among the world’s research centers. I.M.~Khamitov was
supported by the subsidy provided to the Kazan Federal University to
carry out a State assignment in the area of scientific activity.


\begin{thebibliography}{10}
  
\bibitem[\protect\citeauthoryear{Afanasyev and Moiseev}{2005}]{scorpio_05}
  \reference{Afanasyev and Moiseev}[V.L.~Afanasyev, A.V.~Moiseev]
  {\journal{\astl}{31}{194}{2005}}

\bibitem[\protect\citeauthoryear{Afanasyev and Moiseev}{2011}]{scorpio_11}
  \reference{Afanasyev and Moiseev}[V.L.~Afanasyev, A.V.~Moiseev]
  {\journal{Balt. Astr.}{20}{363-370}{2011}}

\bibitem[\protect\citeauthoryear{Burenin et al.}{2007}]{400d}
  \reference{Burenin et al.}[R.A.~Burenin, A.~Vikhlinin, A.~Hornstrup \etal]
  {\journal{\apjs}{172, i.2}{561-582}{2007}}

\bibitem[\protect\citeauthoryear{Burenin}{2015}]{clwise}
  \reference{Burenin R.A.} {\journal{\astl}{41}{167}{2015}}

\bibitem[\protect\citeauthoryear{Vikhlinin et al.}{1998}]{160d}
  \reference{Vikhlinin et al.}[A.~Vikhlinin, B.R.~McNamara, W.~Forman \etal]
  {\journal{\apj}{502}{558}{1998}}

\bibitem[\protect\citeauthoryear{Vikhlinin et al.}{2009}]{cccp}
  \reference{Vikhlinin et al.}[A.~Vikhlinin, A.V.~Kravtsov, R.A.~Burenin \etal]
  {\journal{\apj}{672, i.2}{1060-1074}{2009}}


\bibitem[\protect\citeauthoryear{Wright et al.}{2010}]{wright10}
  \reference{Wright et al.}[E.L.~Wright, P.R.M.~Eisenhardt,,
  A.K.~Mainzer, M.E.~Ressler, R.M.~Cutri, T.~Jarrett,
  J.D.~Kirkpatrick, D.~Padgett, \etal]
  {\journal{\aj}{140}{1868}{2010}}

\bibitem[\protect\citeauthoryear{Planck Collaboration}{2014a}]{PSZcosm13} 
  \reference{Planck Collaboration}
  [Planck 2013 Results XX: P.A.R.~Ade, N.~Aghanim,C.~Armitage-Caplan \etal] 
  {\journal{\aap}{571}{A20}{2014a}}

\bibitem[\protect\citeauthoryear{Planck Collaboration}{2014b}]{PSZcat13} 
  \reference{Planck Collaboration}
  [Planck 2013 Results XXIX: P.A.R.~Ade, N.~Aghanim, C.~Armitage-Caplan \etal] 
  {\journal{\aap}{571}{A29}{2014b}; arXiv:1303.5089}

\bibitem[\protect\citeauthoryear{Planck Collaboration}{2015a}]{Pl_RTT150} \reference{Planck Collaboration}
  [Planck Intemediate Results XXVI: P.A.R.~Ade, N.~Aghanim, M.~Arnaud
  \etal]{ \aap, in press (2015a); arXiv:1407.6663}

\bibitem[\protect\citeauthoryear{Planck Collaboration}{2015b}]{PSZ1Addendum} 
  \reference{Planck Collaboration}
  [Planck 2013 Results XXXII: P.A.R.~Ade, N.~Aghanim, C.~Armitage-Caplan \etal]{
  \journal{\aap}{581}{A14}{2015b}; arXiv:1502.00543}
  
\bibitem[\protect\citeauthoryear{Planck Collaboration}{2015c}]{PSZ2cosm} \reference{Planck Collaboration}
  [Planck 2015 Results XXIV: P.A.R.~Ade, N.~Aghanim, M.~Arnaud \etal]{
    \aap, in press (2015c); arXiv:1502.01597}
  
\bibitem[\protect\citeauthoryear{Planck Collaboration}{2015d}]{PSZ2}
  \reference{Planck Collaboration} [Planck 2015 Results XXVII:
  P.A.R.~Ade, N.~Aghanim, M.~Arnaud \etal]{ \aap, in press
    (2015d); arXiv:1502.01598}

\bibitem[\protect\citeauthoryear {Planck Collaboration}{2015e}]{Canary} \reference{Planck Collaboration} [Planck
  Intermediate Results XXXVI: P.A.R.~Ade, N.~Aghanim, M.~Arnaud
  \etal]{ \aap, in press (2015e); arXiv:1504.04583}
  
\bibitem[\protect\citeauthoryear{Sunyaev and Zeldovich}{1972}]{SZ}
  \reference{Sunyaev and Zeldovich}[R.A.~Sunyaev, Ya.B.~Zeldovich]
  {\journal{\commapsp}{4}{173}{1972}}
\end{thebibliography}
\end{document}